\documentclass{bmcart}

\usepackage{amsthm,amsmath}
\usepackage[utf8]{inputenc} 

\usepackage{graphicx}
\usepackage{subcaption}
	
\usepackage{lineno}
\startlocaldefs
\endlocaldefs

\begin{document}

\begin{frontmatter}

\begin{fmbox}
\dochead{Research}


\title{Author Multidisciplinarity and Disciplinary Roles in Field of Study Networks}


\author[
  addressref={aff1,aff2},                   
  corref={aff1},                       
  email={eoghan.cunningham@ucdconnect.ie}   
]{\inits{E}\fnm{Eoghan} \snm{Cunningham}}
\author[
  addressref={aff1,aff2},
]{\inits{B}\fnm{Barry} \snm{Smyth}}
\author[
  addressref={aff1,aff2},
]{\inits{D}\fnm{Derek} \snm{Greene}}


\address[id=aff1]{
  \orgdiv{School of Computer Science},             
  \orgname{University College Dublin},          
  \cny{Ireland}                                    
}
\address[id=aff2]{%
  \orgdiv{Insight Centre for Data Analytics},
  \orgname{University College Dublin},
  \cny{Ireland}
}



\end{fmbox}


\begin{abstractbox}

\begin{abstract} 
When studying large research corpora, ``distant reading'' methods are vital to understand the topics and trends in the corresponding research space. In particular, given the recognised benefits of multidisciplinary research, it may be important to map schools or communities of diverse research topics, and to understand the multidisciplinary role that topics play within and between these communities. This work proposes Field of Study (FoS) networks as a novel network representation for use in scientometric analysis. We describe the formation of FoS networks, which relate research topics according to the authors who publish in them, from corpora of articles in which fields of study can be identified. FoS networks are particularly useful for the distant reading of large datasets of research papers when analysed through the lens of exploring multidisciplinary science. In an evolving scientific landscape, modular communities in FoS networks offer an alternative categorisation strategy for research topics and sub-disciplines, when compared to traditional prescribed discipline classification schemes. Furthermore, structural role analysis of FoS networks can highlight important characteristics of topics in such communities. To support this, we present two case studies which explore multidisciplinary research in corpora of varying size and scope; namely, 6,323 articles relating to network science research and 4,184,011 articles relating to research on the COVID-19-pandemic.
\end{abstract}




\end{abstractbox}
%

\end{frontmatter}




\section{Introduction}
With the increased recognition of the benefits of multidisciplinary and interdisciplinary collaboration \cite{lariviere_2015,okamura_2019}, a trend has recently been established towards greater levels of interdisciplinary research \cite{leahey_2016}. A common approach for understanding these research processes is through the lens of network analysis. For instance, given a corpus of research papers and their associated metadata, we can construct a variety of  network representations to reveal different aspects of the underlying data, such as co-authorship networks \cite{feng_2020,glanzel04analysing}, citation networks \cite{karunan17discovering}, and co-citation networks \cite{gmur03invisible}. 
These different representations can help us to identify collaboration patterns between individual researchers at a micro level. In other cases we might be more interested in examining collaboration patterns between researchers coming from different disciplines at the macro level. For example, we might wish to study how these patterns evolve over time in response to a changing research funding landscape or impactful exogenous events, such as the COVID-19 pandemic.

In this work, we propose a practical ``distant reading'' approach to help reveal collaboration patterns in large scientific corpora in order to understand better the nature and implications of these patterns. Distant reading has been used in other contexts, such as digital humanities, as a means of exploring large volumes of data from a macro level view, in order to identify specific areas of interest for closer examination \cite{moretti2013distant}. As the core contribution of this work, we present a novel graph representation, referred to as the \emph{Field of Study} (FoS) network, which facilitates the investigation of multidisciplinary and interdisciplinary research in corpora of scientific research articles at the macro level. A key aspect of field of study networks is the use of author-topic relations. Specifically, a FoS network is populated by fields of study (or research topics), which are related to one another according to the authors who publish in them. In Section \ref{sec:methods} we describe how these networks can be constructed from the topics/fields of study that have been assigned to research papers. Later in Section \ref{sec:case_studies} we describe two cases studies, which analyse the FoS networks arising from datasets of differing scope and size. The first case study in Section \ref{sec:ans_case_study} relates to multidisciplinary research in the area of applied network science, while the second study in Section \ref{sec:distant_reading_covid} pertains to the changing nature of author multidisciplinarity in response to the COVID-19 pandemic. These case studies demonstrate that FoS networks can provide a useful tool for the distant reading of large collections of research articles. In particular, we show how simple characteristics computed on a FoS network can highlight important topics in the research corpus. Further, we use community detection methods to identify specific multidisciplinary schools within a larger body of research, and we conduct a `role analysis' of the topics within these communities to understand the role that they play in multidisciplinary collaborations. Crucially, we demonstrate our methods using datasets or varying size and scope, and, finally, we discuss some techniques that may be employed to drill-down on interesting interactions in the graph for further ``close-reading''. 

\section{Related Work}
\label{sec:related_work}
While a range of different definitions exist for \emph{multidisciplinary research}, it is most commonly characterised as work which draws on expertise, data or methodology from two or more distinct disciplines. Most formal definitions distinguish \emph{interdisciplinary} research as an extension of multidisciplinary research, which involves the \emph{integration} of methods from the contributing disciplines \cite{choi2006interdisciplinarity}. 
There are numerous analyses which have explored multi- or interdisciplinary research, and investigated the relationship between different scientific disciplines. 
Many of these studies proposed metrics to quantify research interdisciplinarity, either at the author or at the paper level \cite{rafols2010diversity,porter2007measuring}, often in order to investigate a correlation between interdisciplinarity and research impact \cite{lariviere_2015,okamura_2019}, productivity or visibility \cite{leahey2017prominent}. Typically, works which integrate methods and ideas from a diverse set of disciplines are found to have greater research impact and visibility compared to those that do not \cite{okamura_2019,leahey2017prominent}. Notably, there are several examples of works which have investigated cross-disciplinary collaboration, often drawing on representations and methods from network science \cite{feng_2020,karunan17discovering,wu_2019,raimbault19exploration,lafia21mapping}. 

Most frequently, \emph{co-authorship networks} have been used as a means of representing the collaborations between different researchers, both in small-scale studies and when analysing large-scale bibliographic collections \cite{arnaboldi2016analysis}. 
In this type of network, researchers are represented by nodes and collaborations (i.e., articles jointly authored by a pair of researchers) are encoded by the undirected edges between them. Thus, research teams are identified as fully-connected components of the graph. In cases where research backgrounds can be identified among the authors in the network, this can be used to measure the extent to which authors engage in multidisciplinary collaborations. The analysis of co-authorship networks has often revealed a strong disciplinary homophily between researchers, despite the fact that those with diverse neighbourhoods in these networks tend to have a higher level of research impact \cite{feng_2020}.

Another common representation used to investigate interdisciplinary research is the \emph{citation network}, which is typically constructed at the article or journal level \cite{newman2018networks}. Analyses of citation networks can highlight influential or ``disruptive'' articles in interdisciplinary research \cite{wu_2019}, as well as ``boundary'' papers which span multiple disciplines \cite{karunan17discovering}. Indeed community finding approaches have been employed to automatically group articles in citation networks into their respective fields of study \cite{raimbault19exploration}, so that interdisciplinary interactions can then be explored at the macro level. 

An alternative strategy for analysing research collections is to apply text mining to article abstracts or full-texts in order to group articles together which relate to similar research themes, using techniques such as document clustering or topic modelling \cite{raimbault19exploration,lafia21mapping,yau2014clustering}. This is typically based on word co-occurrence patterns, rather than based on article citation patterns. Of course, the patterns which emerge from textual analysis can be quite different from those generated using network-based approaches, as fields of study which are distant in their authorship or citation representations may still potentially be closely linked semantically. 

Here we propose an alternative network representation, which relates fields of study according to the authors who typically publish in those fields. This kind of network may be used in conjunction with more conventional network representations --- in much the same way that semantic networks have been shown to complement citation networks \cite{raimbault19exploration}. However, later in Section \ref{sec:distant_reading_covid} we show that, on their own, FoS networks can provide an effective means of exploring large scientific collections, particularly in revealing aspects around author multidisciplinarity. 

\section{Methods}
\label{sec:methods}
In this section we formalise the definition of a Field of Study (FoS) network and explain how such a network can be generated from existing research resources. In Sections \ref{sec:static} and \ref{sec:temporal} we describe two FoS variations: the \emph{static} FoS network and the \emph{temporal} FoS network respectively.

\subsection{Field of Study Networks}
Formally, a \emph{Field of Study} network is defined as a general graph representation of a collection of research articles ($R$), written by a set of authors ($A$), and denoted $F = (N, E)$.
The nodes ($N$) represent identifiable research topics (i.e. the fields of study) and the edges ($E$) represent authorship relations between pairs of topics. These relations are aggregated across multiple associated research papers. Below we describe how a FoS network can be constructed from a more conventional authorship graph and we argue that FoS networks are particularly well-suited to analysing the nature of collaboration within the scientific literature, especially as they relate scientific fields of study according to the researchers/authors who publish in them.

The formation of a FoS network depends on the availability of appropriate fields of study labels for a given set of research papers. These could be derived via manual annotations by domain experts, the application of automated text mining methods, or some combination of the two. For instance, topic modelling techniques have been shown to be successful in extracting research topics from corpora of research articles and assigning papers to those fields \cite{lafia21mapping,paul09topic}. 

In fact, many research databases and search engines employ these techniques (or manual classification) to assign articles or academic journals to fields of study. For example, the Microsoft Academic Graph (MAG)\footnote{https://www.microsoft.com/en-us/research/project/microsoft-academic-graph/} maintains a deep hierarchy of \emph{Fields of Study} which they assign to papers; Web of Science (WOS)\footnote{https://clarivate.com/webofsciencegroup/solutions/web-of-science/} group journals in 258 \emph{Subject Categories};  Scopus\footnote{https://www.scopus.com/home.uri} employs experts to assign \emph{All Science Journal Classification (ASJC)} codes to all journals covered by their index. 
For the purpose of the case studies described later in Section \ref{sec:case_studies}, we use MAG fields of study to categorise research papers and construct FoS networks. The deep MAG field of study hierarchy is desirable as it supports the construction of FoS networks at varying levels of detail, from the broadest research disciplines (level 0) to the specific topics and sub-topics that exist within a particular discipline (levels 4 and 5).

It is important to note that the Microsoft Academic Graph may not always be an appropriate source for field of study data. For instance, the corpus does not provide full coverage of all research disciplines and the corresponding hierarchy of fields may contain some spurious connections due to its size and the semi-automated nature of its construction. However, the methods that we propose are not specific to the MAG hierarchy. Rather, they are agnostic in the sense that they are designed to generalise to any case where fields of study can be identified at an appropriate level of detail.

\subsection{Static FoS Networks}
\label{sec:static}

The formation of a static FoS network from a collection of research articles is best described as the two-step process illustrated in Figure \ref{fig:static}. In the first step, an unweighted bipartite graph is generated from identifiable fields of study and their contributing authors; see Figure \ref{fig:static_a}. In the second step, this graph is used to generate a projection (the FoS Network) in which a weighted undirected edge exists between two fields if and only if at least one author has published research in both fields; see Equation \ref{eq:edge} for all $a \in A$, where $N$ is the set of fields identifiable in $R$. The resulting edge weights correspond to the number of such authors who publish in both fields (Equation \ref{eq:weight}). 
\begin{equation}
    E = \big\{(n_i, n_j)~:~published(a, n_i)~\land~published(a, n_j)\big\}
    \label{eq:edge}
\end{equation}
\begin{equation}
    w\big(n_i, n_j\big) = |\big\{a~:~published(a, n_i)~\land~published(a, n_j)\big\}|
    \label{eq:weight}
\end{equation}

\subsection{Temporal FoS Networks}
\label{sec:temporal}
It is further possible to encode temporal information in a FoS Network as \emph{directed} edges, which allows us to study changes in multidisciplinarity research patterns over time. Temporal FoS networks can be visualised in a time-unfolded representation, where the data is divided into a sequence of two or more discrete \emph{time steps}, as frequently employed in dynamic network analysis tasks. 
Nodes are duplicated for each time step so that authors can be connected to any fields in which they publish research during a given time step. 

As an example, Figures \ref{fig:temporal_a} and \ref{fig:temporal_b} illustrate the two stages in the formation of a temporal FoS network, showing an instance of a temporal FoS network with respect to two time-points ($t_n$ and $t_{n+1}$) on either side of some event ($e$); thus $t_n<t_e<t_{n+1}$).
The temporal FoS network in Figure \ref{fig:temporal_b} contains a directed edge between two fields $(n_i,n_j)$ if an author published in field $n_i$ at time $t_n$ (before event $e$) and in field $n_j$ at time $t_{n+1}$ (after event $e$), as given by
\begin{equation}
    E' = \big\{(n_i, n_j)~:~published(a, n_i,t_n)~\land~published(a, n_j,t_{n+1})\big\}
    \label{eq:driected_edge}
\end{equation}
In the next section we present two illustrative examples which demonstrate the utility of static and temporal FoS representations, as described above. 

\section{Case Studies}
\label{sec:case_studies}
In our first case study, presented in Section \ref{sec:ans_case_study}, we consider the use of static FoS networks to explore aspects of multidisciplinary research in the area of network science. The second case study, described in Section \ref{sec:distant_reading_covid}, considers the use of both static and temporal FoS networks in the context of a large-scale dataset of research publications relating to the COVID-19 pandemic.

\subsection{Multidisciplinary Research in Network Science}
\label{sec:ans_case_study}
{\textbf{Network construction.}} Firstly, we focus on research published in the journal Applied Network Science (ANS)\footnote{https://appliednetsci.springeropen.com}, to use as a smaller case study with which we can highlight our methods. We choose ANS as it is a journal with multidisciplinary implications, and we consider the year 2019 as the period with the best coverage in our data source. 
Figure \ref{fig:applied_network_science} presents two resulting static FoS networks, which we create to explore author multidisciplinarity in our data.
These networks are produced using Microsoft Academic Graph metadata for 6,323 research articles. This set of articles represents 131 papers published in the journal Applied Network Science, supplemented by any additional research published by the same authors in the three years prior (2016-2018 inclusive). We use MAG \emph{fields of study} metadata to categorise these research papers. The MAG uses hierarchical topic modelling to identify and assign research topics to individual papers, each of which represents a specific field of study \cite{shen_2018}. 
To produce a more useful categorisation of articles, we consider only those topics at the first two levels of the MAG hierarchy: 
\vskip 0.1em
\begin{enumerate}
    \item The 19 field labels at level 0, which we refer to as `disciplines’.
    \item The 292 field labels at level 1, which we refer to as `sub-disciplines’
\end{enumerate}
\vskip 0.2em
Thus, each article is associated with at least one discipline (e.g. `Medicine’, `Physics’, `Engineering’) and at least one sub-discipline (e.g. `Virology’, `Particle Physics’, `Electronic Engineering’). Note some MAG sub-disciplines belong to more than one discipline. For example, `Biochemistry’ is a child of both `Chemistry’ and `Biology’. 

To center the FoS networks in Applied Network Science research, we include only those edges that originate from ANS papers. We apply weight thresholding to represent the FoS network as unweighted graphs. All analysis is completed on the unweighted graph produced with threshold 5 (the mean edge weight in the weighted network). In order to provide the clearest visualisations, we further prune the networks with threshold 10 before plotting. Figure \ref{fig:ans_a} illustrates the resulting FoS network when network science articles are categorised at the \emph{discipline} level. Each node (or discipline) in this FoS network can then be decomposed into its \emph{sub-disciplines}, as shown in Figure \ref{fig:ans_b}. 

\vskip 0.5em
\noindent{\textbf{Network characterisation.}} From Figure \ref{fig:applied_network_science}, we can begin to understand the multidisciplinarity of authors publishing in Applied Network Science, as the nodes represent a diverse set of sub-disciplines, coloured according to their parent-disciplines. Highly central in Figure \ref{fig:ans_b} are the fields which represent the technical and methodological foundations of network science research. Sub-disciplines of Mathematics and Computer Science, such as `Theoretical Computer Science' and `Topology', have high degree centrality (ranked 1st and 4th respectively), because they are identified across the majority of network science research papers. Modern network science methods, such as `Artificial Intelligence', `Machine Learning' and applications, such as `Information Retrieval', have similarly high degree centrality (ranked 2nd, 3rd, and 6th respectively). Some fields beyond the disciplines of Computer Science and Mathematics, such as `Applied Psychology', `Econometrics', and `Neuroscience' have high betweenness centrality in the FoS Network (ranked 3rd, 5th and 8th, respectively). This is likely because they represent interdisciplinary applications of network science published by authors who have backgrounds in other, more distant topics. For example, in the bottom of Figure \ref{fig:ans_b} we can see a group of medical fields which are linked to topics in Mathematics and Computer Science through `Applied Psychology' and `Social Psychology'.

\vskip 0.5em
\noindent{\textbf{Community detection.}} The MAG FoS hierarchy offers one possible definition of science's traditional disciplinary taxonomy, grouping fields (or sub-disciplines) into broader schools of research. We can explore an alternative categorisation of the topics in the ANS graph by employing community detection methods. Figure \ref{fig:clustering} shows the network from Figure \ref{fig:ans_b}, but with the nodes colour-coded to show cluster memberships identified using the Louvain method \cite{blondel2008louvain} (with resolution parameter value 1.0). This technique identified 7 clusters which maximise modularity in the graph, and group topics according to authorship relations. Table \ref{tab:ans_louvain} provides descriptive statistics for the communities. Such communities represent multidisciplinary clusters of fields across which authors -- in particular, those authors who contributed to ANS in 2019-- are likely to publish. Louvain found clusters containing as few as 2, and as many as 26 topics. Broadly, the clusters can be categorised as: (i) central applied network science topics and applications (ii) networks in machine learning and neuroscience, (iii) psychology, biology and medicine, (iv) mathematics, statistics and natural language processing, (v) product development and process management, (vi) physics and economics, (vii) transport networks and microeconomics.

\vskip 0.5em
\noindent{\textbf{Role analysis.}} In addition to categorising ANS-related topics according to (i) a traditional disciplinary hierarchy (Figure \ref{fig:ans_b}), and (ii) author-related communities (Figure \ref{fig:clustering}), it is also possible to group fields according to the ``role'' they play within the Field of Study network. Using the popular \emph{struc2vec} algorithm \cite{ribeiro2017struc2vec}, we learn dense vector representations for the fields in the FoS network, which preserve \emph{structural equivalence} between nodes. That is, nodes having similar structural features in the graph will have similar representations (commonly known as their role embedding) \cite{rossi2020proximity}. We then cluster the embedding space to identify a discrete set of disciplinary roles. Figure \ref{fig:roles} illustrates the role assignments learned in the ANS graph according to a $k$-means clustering ($k=9$) of struc2vec role embeddings, where $k=9$ represents the elbow of the curve when silhouette scores are plotted for clusterings of increasing values of $k$. Table \ref{tab:roles_ans} shows the mean network centrality scores computed for the different clusters such that we can explain the roles that they represent. Fields in cluster \#1 exhibit ``hub-like'' behaviour, as they score highly for all centrality measures. For each of the largest Louvain communities (i.e. excluding communities (v) and (vii)), the most central node was assigned to role \#1. We will refer to these as the ``core'' nodes since they represent the fields most commonly identified in ANS research and are the most central in the FoS graph. Clusters \#6, \#7, \#8 and \#9 all represent peripheral/leaf nodes with degree 1 and very low centrality scores. None of the topics in the peripheral clusters can be identified in ANS published research. Instead, these topics appear in the 2016--2018 portion of the data and we refer to them as  ``distant background'' topics.

Clusters \#5, \#4 and \#3 are made up of increasingly prevalent background topics. Similar to the distant background roles, a majority of the topics in these clusters never appear in ANS research published in 2019. However, with greater degree than the more peripheral nodes, topics in clusters \#5, \#4 and \#3 appear more frequently in author backgrounds. In the particular case of cluster \#3, we identify a set of ``ANS-adjacent'' disciplines, i.e. the fields in which ANS authors publish the most readily. Finally, cluster \#2 includes non-core topics that have high degree and betweenness centrality. The set of 9 fields in this cluster are separate to the dense communities at the core of the graph. Instead topics like `Applied Psychology', `Computational Biology' and `Regional Science' link distant background subjects to the rest of the network. With all fields in cluster \#2 represented in ANS research published in 2019, we anticipate that the research assigned to these topics offer multidisciplinary applications of network science research, published by authors with diverse research backgrounds. The roles identified in clusters \#1, \#2 and \#6 are apparent in clusterings with 5 $\leq k \leq$ 10 (i.e., identical clusters are found for those parameter values).

 \subsection{Author Multidisciplinarity in COVID-19 Research}
\label{sec:distant_reading_covid}
Field of Study networks generated on yearly data snapshots have been implemented to quantify author multidisciplinarity, according to the extent to which authors publish across different disciplines \cite{cunningham2021collaboration,cunningham2021fos}. They show a stable trend with author multidisciplinarity increasing year-on-year, with a much larger than expected increase for COVID-19-related research. In particular, these analyses grouped research topics (sub-disciplines) according to the MAG disciplinary hierarchy. In the following case study, we explore richer groupings of COVID-19 related research topics in an FoS network, to identify modular communities of sub-disciplines, and to explore their disciplinary roles.

\vskip 0.5em
\noindent{\textbf{Network construction.}} Using a large dataset of COVID-19 related research --  COVID-19 Open Research Dataset (CORD-19)\footnote{https://www.semanticscholar.org/cord19} -- we identify all authors who published COVID-19 related research in 2020, and collect MAG metadata for their COVID-19-related articles, along with any available articles that they published between 2016 and 2019, inclusive. This result is 4,184,011 articles, with 166,356 related to COVID-19. We then construct a FoS network using MAG sub-disciplines identified in the papers. Similar to the ANS example in Section \ref{sec:ans_case_study}, we consider the graph induced by only those edges which originate in COVID-19 research. That is, we do not consider authorship relations between the topics in the pre-COVID-19 portion of the data (2016-2019).
Again, we apply thresholding to produce an unweighted graph where edges with weight greater than or equal to the mean edge weight (50) are preserved. 

\vskip 0.5em
\noindent{\textbf{Community detection.}} When applied to the COVID-19-related FoS network, the Louvain \cite{blondel2008louvain} method (with resolution 1.0) identifies 7 communities, leaving 42 nodes unassigned to any community. Summary statistics for these communities are provided in Table \ref{tab:covid_louvain}. Community (i) groups the core topics in Medicine. It is a dense community with many authors  publishing across almost all pairs of topics. `Surgery', `Pathology' and `Radiology' are the most central fields. Community (ii) is more multidisciplinary than community (i). In addition to many medical fields (`Intensive Care Medicine', `Emergency Medicine', etc.), it contains a number of sub-disciplines in Engineering (e.g. `Engineering Management' and `Electrical Engineering'). As such, the authors who link topics in this community may represent those who tackled the medical emergency posed by the pandemic and, in particular, the challenges associated with the massive strain on intensive care units and relevant equipment like ventilators. 
Community (iii) clearly demarcates those topics relevant to the study of the socioeconomic implications of the pandemic. In addition to topics in Economics, this community links many sub-disciplines of Business and Sociology (e.g. `Financial Systems' and `Demography').

Topics in Biology, Chemistry, Physics and Engineering are linked in community (iv). As the largest and least dense of the communities, (iv) represents the many STEM research areas that are relevant to the study of epidemiology. `Virology', `Immunology', `Computational Biology' and Pharmacology' are among the most central sub-disciplines in community (iv). Community (v) contains topics relevant to Machine Learning and Mathematics and is likely formed as a result of the sizeable effort to apply machine learning and data science methods to detecting and tracking the spread of COVID-19 \cite{nguyen_2021}. Finally, communities (vi) and (vii) represent the smallest and most dense communities in the FoS network. 

The topics in community (vi) relate to studies of the environmental impact of the COVID-19 pandemic and associated lockdowns, while nodes in community (vii) are related to `Astrophysics'. Further inspection of the sub-disciplines in community (vii) (`Astrophysics', `Astronomy', `Classical Mechanics', and `Computer Engineering') highlights a portion of the CORD-19 dataset that is unrelated to the COVID-19 pandemic. We believe these papers were included in the collection in error. The modular FoS communities represent groups of topics which are strongly related according to the authors who publish in them. As such, these communities highlight the different schools/disciplines which emerged in COVID-19 research, together with the different research backgrounds and expertise with which authors contributed to them. Crucially, these disciplines offer an alternative classification of sub-disciplines to the more traditional MAG scheme, highlighting instead a more nuanced, multidisciplinary set of topics, specific to the pandemic. 

\vskip 0.5em
\noindent{\textbf{Role analysis.}} We also conduct a role analysis of the topics in the COVID-19-related FoS network, using the methods described in the ANS case study above. As before, we identify a discrete set of roles via k-means clustering of struc2vec role embeddings. We consider an optimal clustering to be the elbow of the silhouette score curve when plotted for increasing values of $k$. Consistent with the greater scope of the COVID-19-related dataset (when compared with that of the ANS dataset), we identify a larger set of clusters in the COVID-19-related FoS network ($k$ = 21). Statistics for these clusters are provided in Table \ref{tab:roles_covid}. Although the clusters appear more numerous and complex than in the ANS case study, a number of distinct roles are evident. We now discuss the predominant roles in turn.

The disciplinary hubs in the graph are captured in role \#1. `Internal Medicine', `Environmental Health', `Virology', and `Artificial Intelligence' are clustered in role \#1 as the core nodes in the network, with each topic among the most central nodes in the Louvain communities (ii), (iii), (iv) and (iv) respectively. Topics in role \#4 have high betweenness scores, despite being outside of the most the central core of the graph (according to eigenvector centrality).  Similar to role \#2 in the ANS case study, these topics likely play a bridging role, linking otherwise disconnected topics to the rest of the graph. Role \#4 contains topics such as `Economic Growth', `Algorithm', `Social Psychology' and `Risk Analysis', which all fall outside of the scope of virology or epidemiology. These topics occur in COVID-19-related research that is published by authors with research backgrounds that are more peripheral in the graph. We hypothesise that topics attributed to this bridging role occur in multidisciplinary applications of one or more fields to a external problem. A similar bridging role may be described by role \#9, which has high betweenness centrality (ranked 4th), but relatively low eigenvector centrality (ranked 9th). With lower eigenvector centrality, it is unlikely that nodes in role \#9 are adjacent to highly central topics in the graph and, as such, likely represent more peripheral ``bridging'' disciplines. Topics in role \#9 are `Composite Materials', `Computer Network', `Atmospheric Sciences' and `Climatology'. The largest cluster in the graph is role \#15. With relatively low degree (mean = 3.2, median = 3) and greater eigenvector centrality than nodes with similarly low degree, it is likely that this cluster represents background topics which are adjacent to two or more of the more central COVID-19-related topics. Although the topics in this role are quite diverse, the cluster contains many sub-disciplines of Engineering, Chemistry, and Physics.

Through the role analysis outlined above, it is possible to further categorise the topics in the FoS graph according to the role they play within and between the disciplines described by the Louvain clusters. In particular, we identify those topics that (i) are at the core of the discipline(s) (i.e., hubs), (ii) represent multidisciplinary applications (i.e., bridges), (iii) are relevant in the research backgrounds of contributing authors (i.e., leaf or peripheral nodes).

\vskip 0.5em
\noindent{\textbf{Close reading.}} For very large datasets, such as the COVID-19-related research explored in this case study, it can be difficult to parse FoS network visualisations. Thus, we rely on computational methods such as community detection and role analysis to understand the relationship between fields of study. Such methods describe the network structure and the multidisciplinary role of the associated research topics as we have shown. Additionally, these methods can highlight cases of multidisciplinary research which can be explored in greater detail. For example, Figure \ref{fig:covid_louvain_3} presents the FoS subgraph containing the topics in Louvain community (iii). We highlight these topics as they represent one of the larger, more multidisciplinary communities that were identified in COVID-19-related research dataset. This community groups many topics from the disciplines Medicine, Economics, Psychology, Sociology and Political Science. The authors who link these topics likely represent those who contributed research relating to the socioeconomic impact of the pandemic. Highly central in the subgraph are sub-disciplines of Medicine such as `Family Medicine' and `Gerontology' (the study of the social, psychological and biological aspects of ageing), in addition to non-Medical topics like `Economic Growth' and `Demography' (the statistical study of populations).

Many topics from Psychology and Economics are present in the more peripheral nodes in the graph, as are sub-disciplines of Mathematics and Computer Science (e.g. `Internet Privacy' and `Statistics') and even topics from Political Science (e.g. `Public Relations' and `Public Administration'). The FoS subgraph helps to illustrate the highly diverse school of research that developed around the study of the socioeconomic impacts of the pandemic. We can further investigate the multidisciplinary nature of the research in this subset by using Temporal FoS networks to compare the pre-COVID (2016--2019) and COVID (COVID-19 related research in 2020) time periods. For example, we might ask the question -- `What were the research backgrounds/expertise of the authors who published COVID-19-related research in the field of Economics?'. 

Figure \ref{fig:flow_economics} presents COVID-19 related research in the field of Economics, with pre-COVID nodes on the left (representing the authors' research backgrounds) and COVID nodes on the right (representing the FoS characterisation of the COVID related research). To highlight the strongest trends that exist, the FoS network shows only the top-30 edges by weight prior to thresholding. The multidisciplinary nature of this research subset is apparent in the diverse set of topics illustrated on the left hand side of the plot. In accordance with the broad spectrum of factors (social, political and economic) which influenced economic growth during the pandemic, we identify many authors who have published previously in sociology, psychology and political science in the graph. Additionally, those topics which may have useful, transferable skills such as `Statistics' and `Data Science' are also found to contribute.  

To conduct further close reading, we can filter the list of articles by considering only those papers that contribute a particular edge to the FoS network. For example, we can search for COVID-related papers which result in the edge between `Social Psychology' and `Economic Growth'. These will correspond to COVID-related articles containing the topic `Economic Growth', in which the authors have previously published research in the field of social psychology. To better understand the papers in this subset, we can explore the lower-level MAG topics that are most commonly identified amongst them, or the keywords which occur most frequently in their titles and abstracts.

\section{Conclusions}

In this work we have demonstrated that our proposed Field of Study (FoS) networks provide a useful means of exploring author multidisciplinarity in a body of research. The two case studies, provided in Sections \ref{sec:ans_case_study} and \ref{sec:distant_reading_covid}, have shown the utility of FoS networks for this purpose in mid- ($\approx 6,000$) and large-sized ($\approx 5,000,000$) research corpora. Modular communities in FoS networks offer an alternative categorisation strategy for research topics and sub-disciplines, when compared to traditional prescribed discipline classification schemes. Such communities represent the broader, multidisciplinary trends in a body of research, together the different backgrounds and expertise with which authors contribute to them. Furthermore, role analysis, using methods such as struc2vec role embeddings, can be employed to parse the respective roles of topics within and between these communities. In particular, we have highlighted core and background roles, which serves to distinguish the central topics in a field from the background expertise of the authors. In addition, less central topics with high betweenness centrality may highlight multidisciplinary applications in the body of research. In the case of very large corpora, visualising FoS networks can be challenging. As such, in Section \ref{sec:distant_reading_covid} we have outlined methods for drilling down to conduct closer reading of research corpora, at greater detail, using dynamic FoS networks. 

There are a number of avenues for potential further research in this area. For example, in a corpus where full paper texts or abstracts are available, it may be informative to explore semantic relationships between the fields of study represented in the network. Similarly, citation information could be used to explore the flow or diffusion of information between communities. Recent work in \cite{raimbault19exploration} suggests that a multi-dimensional approach, which combines these methods, may prove to be a useful tool for scientometric analysis. 


\begin{backmatter}
\section*{Declarations}

\subsection*{Funding}
This research was supported by Science Foundation Ireland (SFI) under Grant Number SFI/12/RC/2289\_P2. 


\subsection*{Consent for Publication}
Not applicable. 

\subsection*{Availability of data and materials}
An archive of the relevant metadata for both case studies will be made available online before any final publication.

\subsection*{Ethics approval and consent to participate}
Not applicable.

\subsection*{Competing interests}
The authors declare that they have no competing interests.

\subsection*{Authors' contributions}
EC proposed the methods, implemented the code, and performed the experiments. EC, DG and BS wrote, read, and approved the manuscript.

\subsection*{Acknowledgements}
Not applicable.





\bibliographystyle{bmc-mathphys} 
\bibliography{references}      




\clearpage
\section*{Figures}

\begin{figure}[!h]
    \centering
    \begin{subfigure}{0.42\linewidth}
        \includegraphics[width=\linewidth]{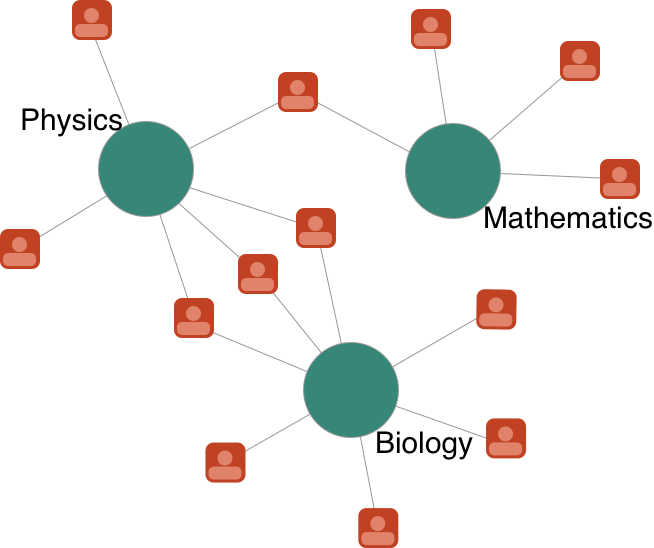}    
        \caption{}
        \label{fig:static_a}
    \end{subfigure}
    \hspace{1.5em}
    \begin{subfigure}{0.43\linewidth}
        \includegraphics[width=\linewidth]{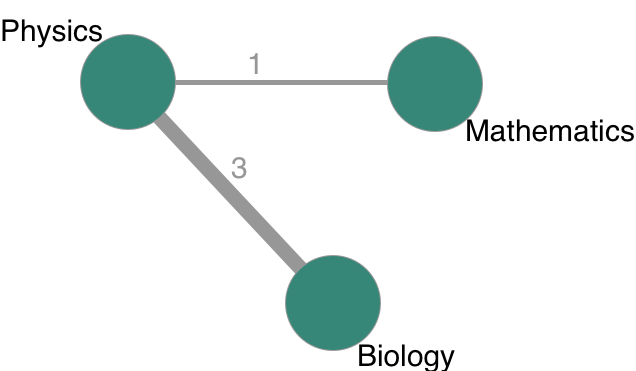}  
        \vskip 0.1em
        \caption{}
        \label{fig:static_b}
    \end{subfigure}
    \caption{The formation of a \emph{static} Field of Study (FoS) network involving two steps: (a)  creation of a bipartite network of authors and fields; (b) projection to an \emph{undirected} network of fields.}
    \label{fig:static}
\end{figure}

\begin{figure}[!h]
    \centering
    \begin{subfigure}{0.41\linewidth}
        \includegraphics[width=\linewidth]{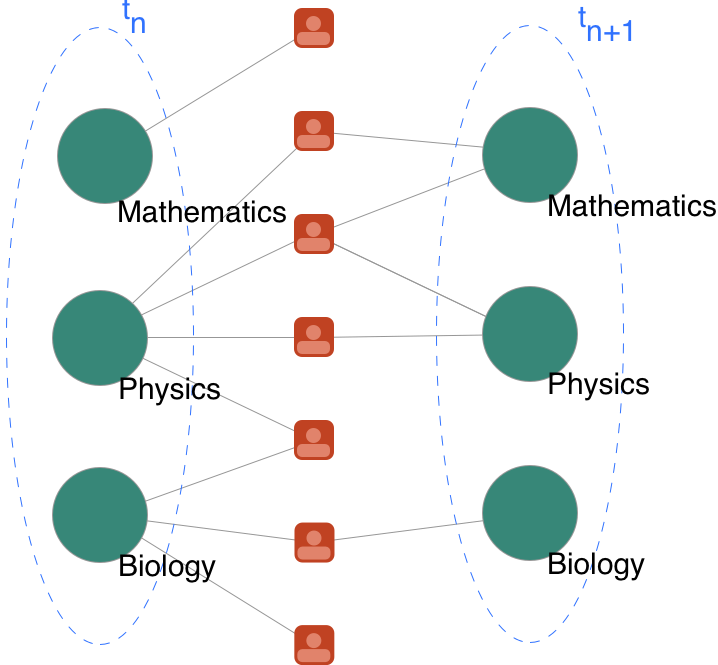}
        \vskip 0.1em
        \caption{}
        \label{fig:temporal_a}
    \end{subfigure}
    \hspace{2.5em}
    \begin{subfigure}{0.41\linewidth}
        \includegraphics[width=\linewidth]{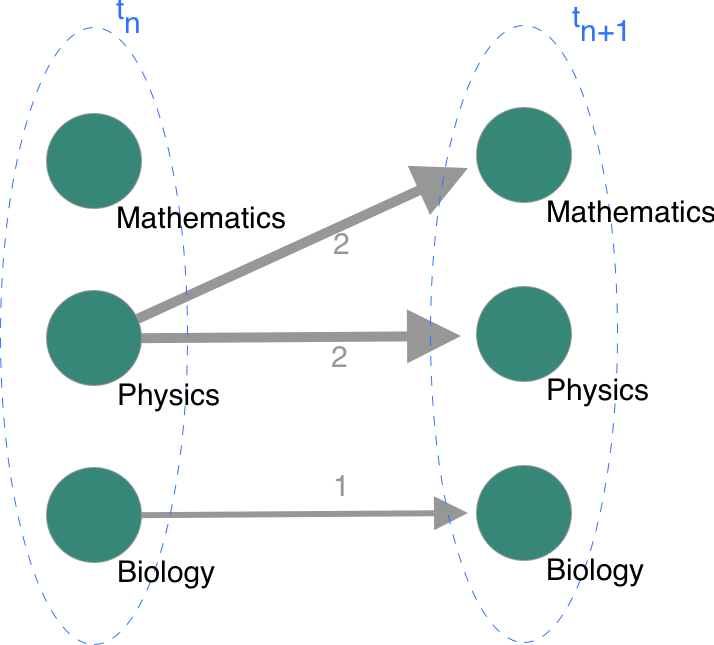}
        \vskip 0.2em
        \caption{}
        \label{fig:temporal_b}
    \end{subfigure}
    \caption{Illustrative example of a \emph{temporal} Field of Study (FoS) network, involving  two steps: (a)  creation of a bipartite network of authors and fields; (b) projection to a \emph{directed} network of fields.}
\end{figure}
  
\begin{figure}[!h]
    \begin{subfigure}{\linewidth}
    \centering
        \includegraphics[width=0.4\linewidth]{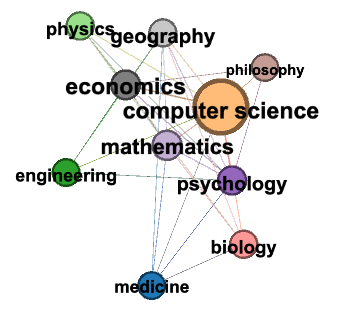}
        \caption{Disciplines or level 0 fields of study.}
        \label{fig:ans_a}
    \end{subfigure}
    \vskip 1em
    \begin{subfigure}{\linewidth}
    \centering
        \includegraphics[width=0.83\linewidth]{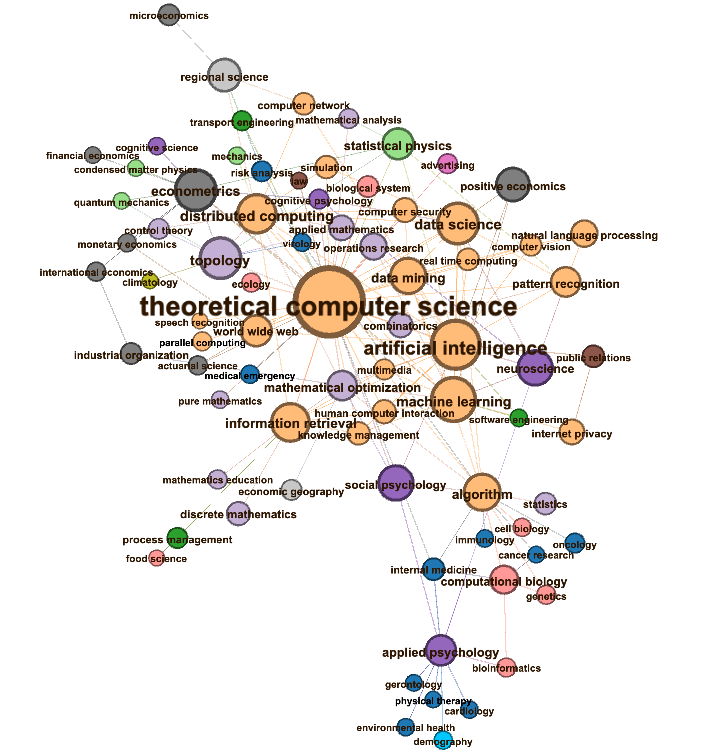}
        \caption{Sub-disciplines or level 1 fields of study.}
        \label{fig:ans_b}
    \end{subfigure}
    \caption{FoS Network for research published in related to the journal ``Applied Network Science'' during 2016--2019. Node size encodes the number of papers attributed to a field of study. In (b) nodes are coloured to represent the parent discipline of the field of study. Edges are coloured to show the parent discipline if the edge is within a discipline/community. Edges between communities are not coloured.} 
    \label{fig:applied_network_science}
\end{figure}

\begin{figure}[!h]
    \centering
    \includegraphics[width=0.83\linewidth]{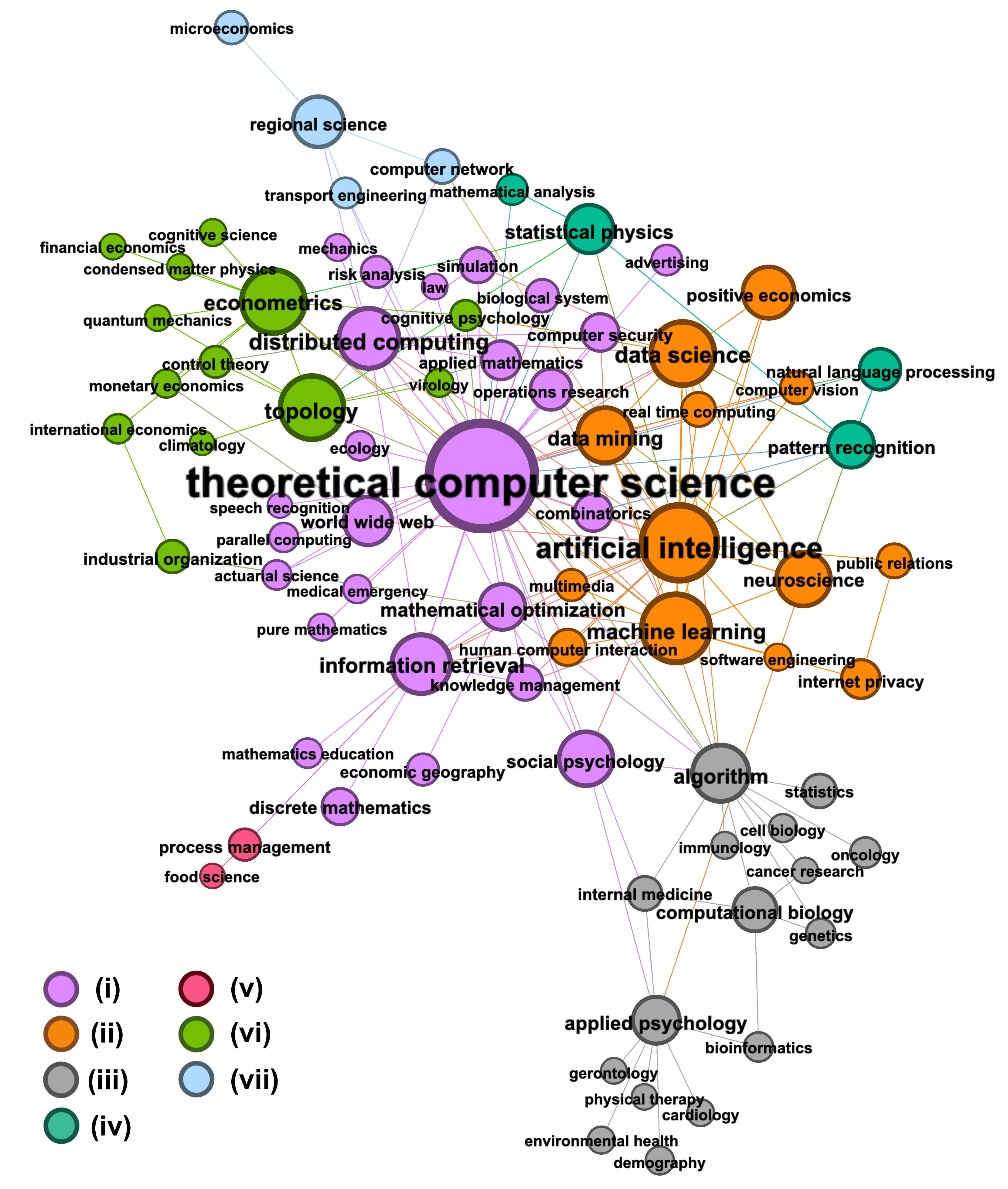}
    \caption{FoS Network for research published in related to the journal ``Applied Network Science'' during 2016--2019. Nodes are coloured to show clusters identified by Louvain.}
    \label{fig:clustering}
\end{figure}

\begin{figure}[!h]
    \centering
    \includegraphics[width=0.83\linewidth]{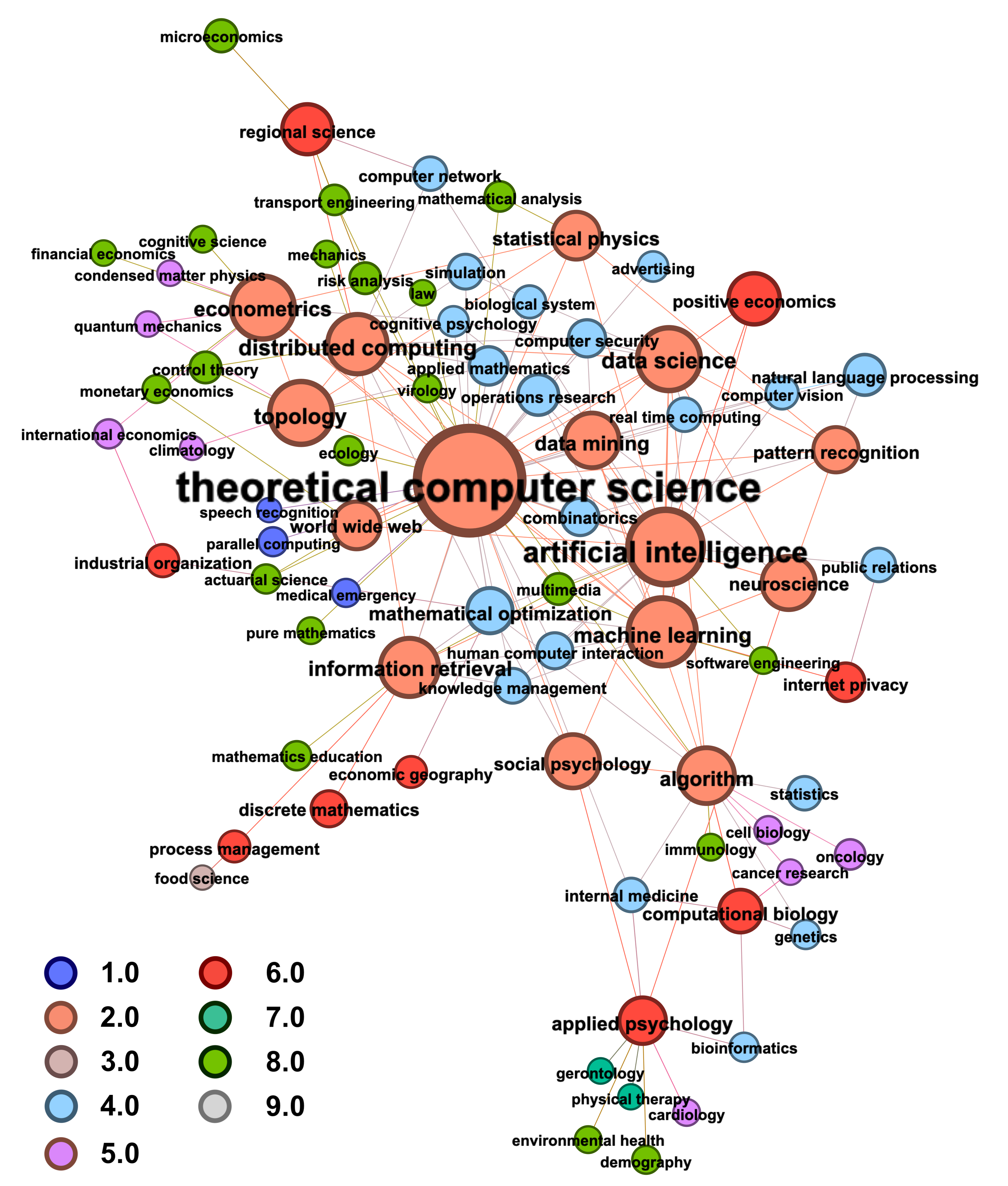}
    \caption{FoS Network for research published in related to the journal ``Applied Network Science'' during 2016--2019. Nodes are coloured to show role assignments according to 9 clusters generated on \emph{struc2vec} embeddings.}
    \label{fig:roles}
\end{figure}

\begin{figure}[!h]
    \centering
    \includegraphics[width=0.83\linewidth]{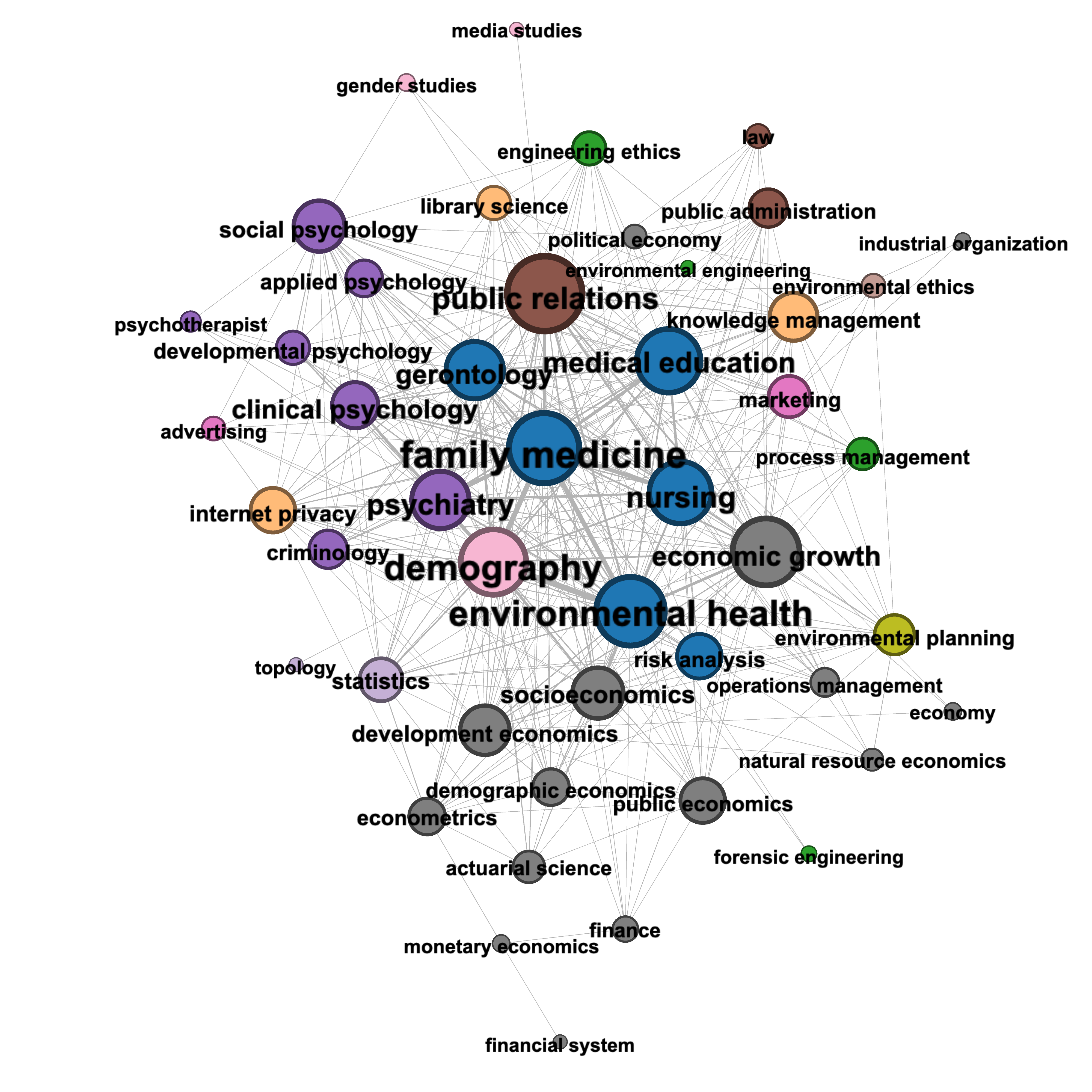}
    \caption{FoS Network for a community of COVID-19-related research published during 2015--2020. Louvain community (iii). Nodes are coloured according to their MAG parent discipline.}
    \label{fig:covid_louvain_3}
\end{figure}

\begin{figure}[!h]
    \includegraphics[width=\linewidth]{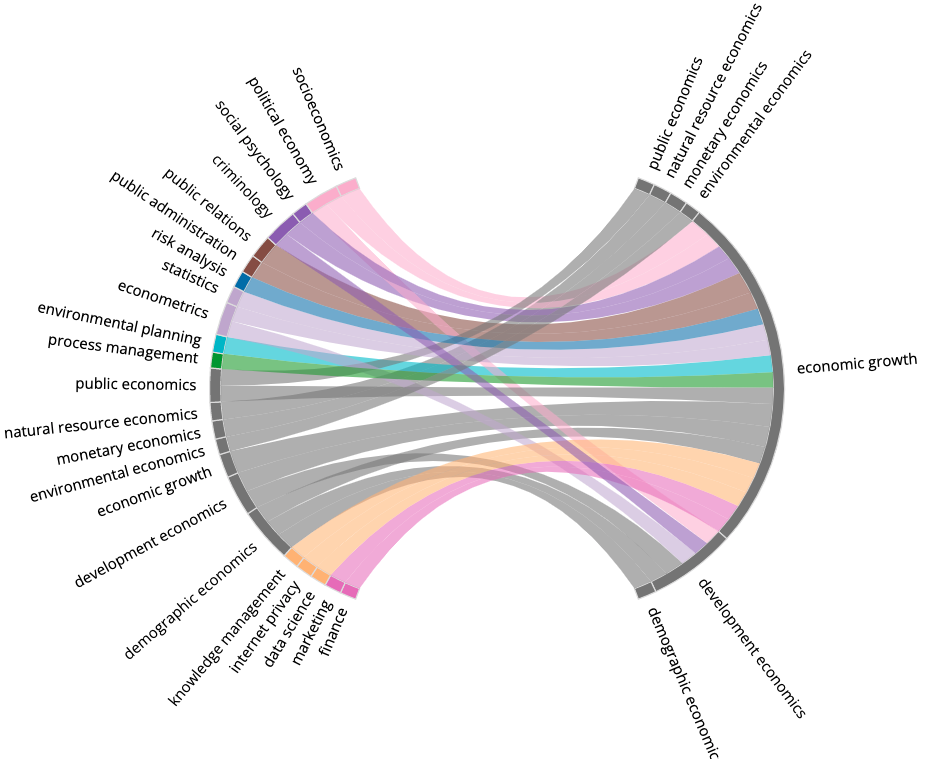}
    \caption{Temporal FoS Network presenting COVID-19-related research in Economics, produced from 1,355 COVID-related research papers which were attributed to the MAG field `Economics'.} 
    \label{fig:flow_economics}
\end{figure}


\clearpage
\section*{Tables}

\begin{table}[!h]
\caption{Louvain communities in ANS-related research and their size, network density, most central topics (according to degree centrality) and the most frequent MAG disciplines that are identified in them ($\geq 20\%$ of topics).}
\label{tab:ans_louvain}
  \begin{tabular}{lrrll}
\hline
{} & size & density &  most central nodes &   MAG disciplines \\
\hline
(i)     &         67 &    0.04 &  theoretical computer science &                     computer science \\
(ii)     &         15 &    0.39 &  artificial intelligence, machine learning &                     computer science \\
(iii)     &         16 &    0.23 &  algorithm, applied psychology &                    medicine, biology \\
(iv)     &          9 &    0.14 &  statistical physics, pattern recognition &  biology, computer science, medicine \\
(v)     &         11 &    0.04 &  process management, food science &                              physics \\
(vi)     &         18 &    0.14 &  econometrics, topology, industrial org &                   economics, physics \\
(vii)     &          5 &    0.40 &  computer network, regional science &                            economics \\
\hline
\end{tabular}
\end{table}

\begin{table}[!h]
\caption{Roles identified in Applied Network Science Research and their mean network attributes, including centrality scores, cluster size (count), and the proportion of topics in the cluster identified in ANS research (ANS prop).}
\label{tab:roles_ans}
  \begin{tabular}{lrrrrrrr}
\hline
role & degree &  betweenness & closeness & eigenvector & ANS prop & count \\
\hline
\#1 & 28.7 & 0.025 & 0.279 & 0.503 & 1.00 & 15 \\
\#2 & 12.2 & 0.014 & 0.223 & 0.171 & 1.00 & 9 \\
\#3 & 7.2 & 0.002 & 0.245 & 0.220 & 0.29 & 21 \\
\#4 & 2.9 & 0.002 & 0.229 & 0.108 & 0.15 & 27 \\
\#5 & 1.7 & 0.000 & 0.187 & 0.038 & 0.10 & 31 \\
\#6 & 1.0 & 0.000 & 0.221 & 0.061 & 0.00 & 25 \\
\#7 & 1.0 & 0.000 & 0.148 & 0.005 & 0.00 & 4 \\
\#8 & 1.0 & 0.000 & 0.159 & 0.011 & 0.17 & 6 \\
\#9 & 1.0 & 0.000 & 0.169 & 0.011 & 0.00 & 3 \\
\hline
\end{tabular}

\end{table}

\begin{table}[!h]
\caption{Louvain communities in COVID-19-related research and their size, network density, most central topics (according to degree centrality) and the most frequent MAG disciplines that are identified in them ($\geq 10\%$ of topics).}
\label{tab:covid_louvain}
  \begin{tabular}{lrrll}
\hline
{} & size & density &  most central nodes &   MAG disciplines \\
\hline
(i)     &         23 &    0.69 &                      surgery, pathology, radiology &                                medicine \\
(ii)     &         22 &    0.26 &  intensive care medicine, emergency medicine &                                medicine \\
(iii)     &         70 &    0.30 &      public relations, economic growth, demography &                               economics \\
(iv)    &         82 &    0.18 &             virology, nanotechnology, cell biology &                      biology\\
(v)     &         42 &    0.27 &   artificial intelligence, algorithm, data science &           computer science, mathematics \\
(vi)     &          5 &    0.90 &  atmospheric sciences, climatology &             geography\\
(vii)     &          4 &    0.83 &       astrophysics, astronomy, classical mechanics &  physics, engineering\\
\hline
\end{tabular}

\end{table}

\begin{table}[!h]
\caption{Roles identified in COVID-19 related research and their mean network attributes, including centrality scores and cluster size (count).}
\label{tab:roles_covid}
  \begin{tabular}{lrrrrrr}
\hline
role & degree &  betweenness & closeness & eigenvector & count \\
\hline
\#\textbf{1} & \textbf{132.2} & \textbf{0.076} & \textbf{0.570} & \textbf{0.951} & \textbf{4 }\\
\#2 & 112.2 & 0.025 & 0.534 & 0.925 & 5\\
\#3 & 84.3 & 0.005 & 0.495 & 0.811 & 29 \\
\#\textbf{4} & \textbf{65.2} & \textbf{0.013} & \textbf{0.474} & \textbf{0.570} & \textbf{12} \\
\#5 & 49.3 & 0.000 & 0.446 & 0.535 & 24 \\
\#6 & 45.5 & 0.003 & 0.437 & 0.418 & 22 \\
\#7 & 19.8 & 0.001 & 0.380 & 0.128 & 26 \\
\#8 & 15.0 & 0.000 & 0.384 & 0.154 & 21 \\
\#\textbf{9} & \textbf{13.5} & \textbf{0.006} & \textbf{0.384} & \textbf{0.074} & \textbf{4 }\\
\#10 & 9.2 & 0.001 & 0.350 & 0.042 & 19 \\
\#11 & 8.2 & 0.000 & 0.303 & 0.016 & 4 \\
\#12 & 5.8 & 0.000 & 0.324 & 0.031 & 6 \\
\#13 & 4.7 & 0.000 & 0.306 & 0.011 & 3 \\
\#14 & 3.7 & 0.000 & 0.300 & 0.011 & 3 \\
\#\textbf{15} & \textbf{3.2} & \textbf{0.000} & \textbf{0.329} & \textbf{0.036} & \textbf{51} \\
\#16 & 2.4 & 0.000 & 0.264 & 0.001 & 7 \\
\#17 & 2.0 & 0.000 & 0.249 & 0.000 & 2 \\
\#18 & 1.0 & 0.000 & 0.306 & 0.007 & 2 \\
\#19 & 1.0 & 0.000 & 0.332 & 0.014 & 5 \\
\#20 & 1.0 & 0.000 & 0.319 & 0.013 & 2 \\
\#21 & 1.0 & 0.000 & 0.323 & 0.013 & 2 \\
\hline
\end{tabular}
\end{table}




\end{backmatter}
\end{document}